\documentclass[prd,superscriptaddress,floatfix,notitlepage,nofootinbib,reprint]{revtex4-1} 
\pdfoutput=1
\usepackage{amssymb,epsfig}
\usepackage{amsmath}
\usepackage[]{empheq}
\usepackage{hyperref}
\usepackage{natbib,ifthen}

\usepackage{wasysym}
\usepackage{mathrsfs}
\usepackage[lofdepth,lotdepth,caption=false]{subfig}
\usepackage{color}
\usepackage{bbold}

\usepackage{mathrsfs}

\usepackage{csvsimple}

\bibpunct{[}{]}{,}{n}{,}{;}

\begin{document}

\newcommand{\jcap}{JCAP}
\newcommand{\apjl}{APJL~}
\newcommand{\aap}{Astronomy \& Astrophysics}
\newcommand{\mnras}{Mon.\ Not.\ R.\ Astron.\ Soc.}
\newcommand{\apjs}{Astrophys.\ J.\ Supp.}

\thispagestyle{empty}
\title{Kepler data analysis: non-Gaussian 
noise and Fourier Gaussian process analysis of star variability}

\author{Jakob Robnik}
\affiliation{University of Ljubljana, Faculty of Mathematics and Physics,\\
Jadranska 19, 1000 Ljubljana, Slovenia}

\author{Uro\v{s} Seljak}
\affiliation{Department of Astronomy and Department of Physics,
University of California, Berkeley, CA 94720, USA}
\affiliation{Lawrence Berkeley National Laboratory, 1 Cyclotron Road, Berkeley, CA
93720, USA}

\date{\today}
\begin{abstract}
We develop a
statistical analysis model of Kepler star flux data
in the presence of planet transits, non-Gaussian 
noise, and star variability.  
We first develop a model for Kepler noise probability distribution  
in the presence of outliers, which make the noise
probability distribution non-Gaussian. We develop a signal likelihood 
analysis based on this probability distribution, in which we model 
the signal as a sum of the star variability and planetary transits. 
We argue these components need to be modeled together if optimal signal
is to be extracted from the data. For the star variability model we 
develop an optimal Gaussian process analysis using a Fourier based 
Wiener filter approach, where
the power spectrum is non-parametric and learned
from the data. We 
develop
high dimensional optimization of the objective function, where   
we
jointly optimize all the model parameters, including thousands of star variability modes, and planet transit parameters. 
We apply the method to Kepler-90 data and show that it 
gives a better match to the star variability than the 
standard spline method, and robustly handles noise outliers. As a consequence, 
the planet radii have a higher value 
than the standard spline method.

\end{abstract}

\maketitle

\section{Introduction}
\label{se:intro}

Kepler Space Telescope operated on its primary mission in the years 2009-2013, aiming to photometrically detect exoplanets using transits. It measured flux from 200,000 stars and detected 18,000 potential planets of which 2,000 are confirmed as of now, spanning the range of masses down to approximately
Earth-sized objects (\cite{smallplanets}). 

A common approach to planet detection is to search through possible periods of planets, folding signal phase wise and seeking high signal to noise events as described in \cite{phase_search}. One of the challenges is to distinguish true planets from false positives. For example, a proposed planet could be an eclipsing binary star, a single or multiple noise event resembling a planet transit, a fluctuation of host star`s brightness, an event in an off-set star, which is a star in the same field of view, but has no physical contact with a given star \cite{false_positive}, etc. A traditional approach is to perform a series of tests, each designed to target a specific group of false positives, and eliminate them if a candidate does not pass these individual tests. Transits are checked for uniformity of transit depths, consistency, possible correlation with other known planets in the given system, and for the shape of transit, by calculating a metric distance (LPP metric) from known planet shapes (\cite{robovetter}).

A first step in the analysis is to have a good probabilistic model of all the 
components that contribute to the observed data. In this paper 
we present such analysis, where we statistically analyze several different 
components constituting the measured flux. Our goal is to develop a probability distribution of the data, which can then be used to asses the 
probability of a given observed transit-like shape to be a planet. Here we will build a model describing the incident flux and analyze its components. We first build a model of noise probability distribution based on the analysis of stars where there are no planets (section \autoref{chap:Noise}). 
It enables us to rigorously analyze outliers and it eliminates the need to use robust statistics such as outlier rejection. 
It has a significant impact on the statistical significance of proposed planets. We proceed by modeling stellar variability and shape of planet transits in section \autoref{chap:Star}. We argue 
it is crucial to analyze them simultaneously. A common practice in literature is to fit spline with different knot spacing, and iteratively removing outliers, while searching for the lowest Bayesian information criterion 
in order to prevent overfitting \cite{spline}. Spline is then removed from the data and planets are fitted separately. This procedure is sub-optimal because it eliminates signal with no knowledge about its origin, potentially removing part of the planet 
signal as well. We propose an algorithm for joint analysis in section \autoref{chap:Joint}, where both star variability and planets are fitted together. We demonstrate use of our analysis in section \autoref{chap:planet}, where we apply it to a signal from Kepler-90, a star known to host eight planets, which is 
the largest known planetary system, together with our own. 

The structure of this paper is as follows. In section 2 we develop the noise probability distribution model. In section 3 we develop 
the flux probability distribution model, accounting for star variabity and planet signal. In section 5 we apply these  
components to analyze planet signals in an example of Kepler-90. This is followed by conclusions in section 6. 

\section{Noise probability distribution model} \label{chap:Noise}

\begin{figure*}[t]
\hspace*{-1.6cm}\includegraphics[scale= 0.45]{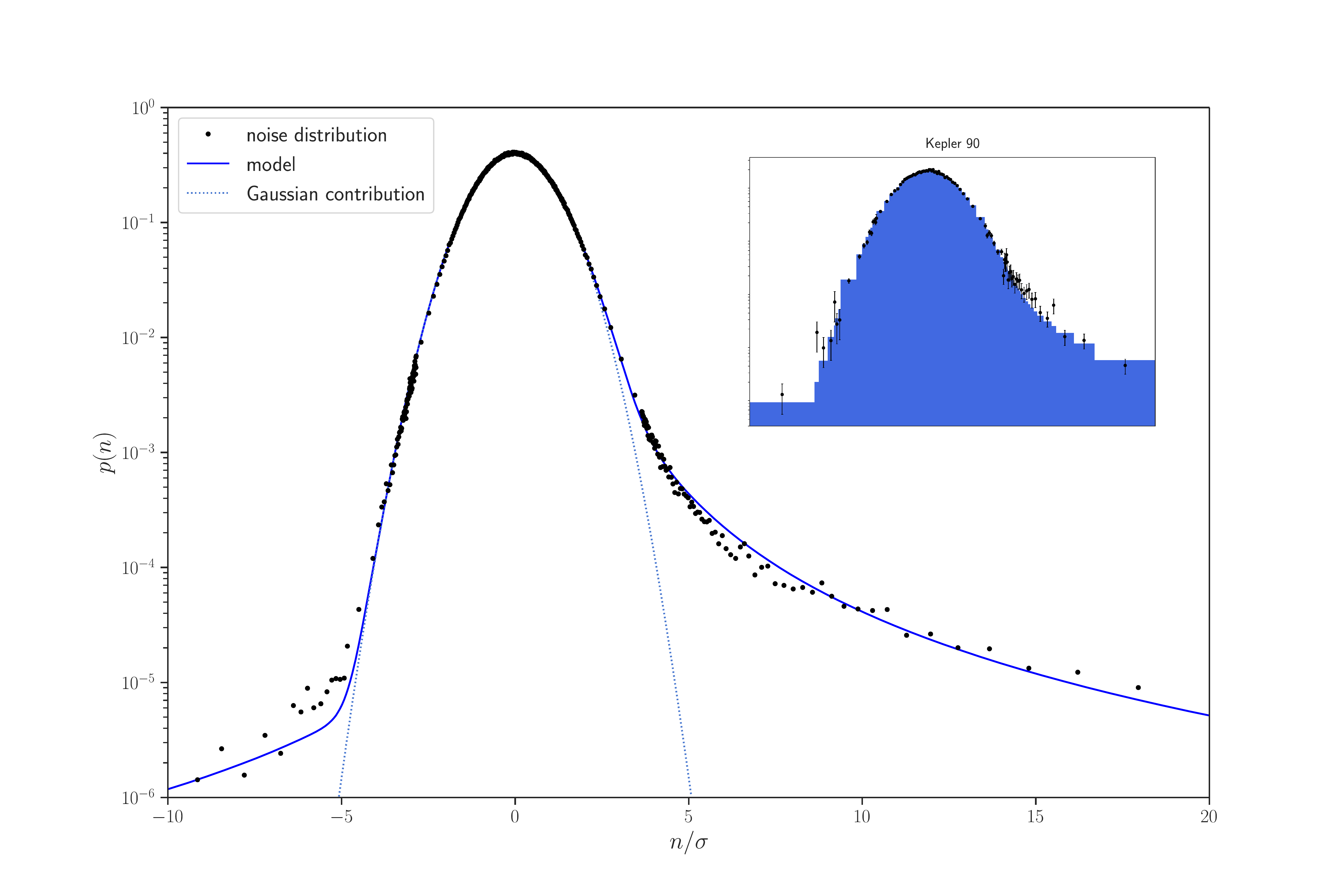}
\caption{Normalized noise probability distribution as a result of combining rescaled flux coming from 76 stars with no planets. Probability density is mostly Gaussian, combined with power law outliers that we model with non central t-distribution. We extract optimal model parameters with minimization of a negative log-likelihood. Assuming that a distribution is Gaussian would result in an underestimation of a probability for outliers. In the insert we show the 
power spectrum estimate from Kepler 90 only, after removing all the known planets. While the errors are larger, the 
result is consistent with the larger dataset of 76 
stars, as shown by the blue model distribution, where 
we interpolated the model to the wider bins of Kepler 90.}
\label{fig:noise_model}
\end{figure*}

In this section we will build a model of a noise probability distribution that will enable us to rigorously analyze planet detections and make use of robust statistics such as outlier removal unnecessary.

We use the Kepler data \footnote{\url{https://exoplanetarchive.ipac.caltech.edu/bulk_data_download/}} processed through the Pre-search data conditioning module \cite{kepler_process}, which eliminates systematic instrumental errors like features of 90 days rotation of a telescope, temperature changes of the aperture etc. We use a so called PDCSAP flux, where long term trends have also been eliminated. For this data reduction noise is observed to be uncorrelated, i.e. flat in the frequency domain (figure \ref{fig:hyperprior}). 

In the absence of planets signal $y(t)$ is composed of star variability, denoted as $s(t)$, and noise $n(t)$:
\begin{equation}
    y(t) =  s(t) + n(t).
\end{equation}
To determine noise distribution we also need to determine 
star variability $s(t)$. We will describe a method to do 
so in next section: here we assume we have this available, 
which means we need to iterate on 
noise and star variability. Assuming we have converged on  $s(t)$,
in figure \ref{fig:noise_model} we show the probability distribution of the 
noise $n(t)$. Majority of the data is normally distributed, except for the outliers. 
A single light curve may not have enough outliers to analyze them statistically, so we will show results from a single 
star after planet removal, as well as 
combining flux from more than one star, picking those stars that are believed to have no planets, which enables us to study the negative outliers as well. 

Flux from different stars has different amplitude and so 
also different level of noise: we rescale it in order to treat it as a realization of the same noise distribution. We choose normalization in such a way that contribution of a Gaussian distribution to the noise has equal variance ($\sigma^2$). To compute $\sigma$ we first identify central, Gaussian part of the distribution. A measurement will be defined to belong to the central part of a distribution if its deviation from the average of a central part is say less than $\sigma$: $|n - \mu| < \sigma$. This choice is arbitrary, it suffices to pick an interval where distribution is still Gaussian, not influenced by outliers. $\mu$ and $\sigma$ are of course unknown and will be obtained by iteration. An initial guess is $\mu_0 = E[n]$ and $\sigma_0^2 = Var[n]$, where $n$ are all noise measurements from a star. In the next step only measurements which are less than current guess on $\sigma$ away from the current guess on an average are taken into account. In this way impact of outliers is eliminated. A new $\sigma$ is computed in such a way to match $Var[ \{ n_i ; \, n_i \in n, \, |n_i - \mu| < \sigma \}] = \sigma^2  \int_{-1}^{1} N(x)^2 \text{d}x$. 
We analyze combined fluxes to extract probability density function. As shown in figure \ref{fig:noise_model}, we observe a combination of Gaussian distribution for small fluxes in the core of pdf and a shallower slope governing the outliers.

We form noise probability density function as a sum of two terms: the
dominant Gaussian contribution, accountimg for the majority of noise events, and a function modeling outliers. Density function for outliers is a power law that stretches to high variance, but becomes irrelevant for small values when compared to the Gaussian contribution. It is also evident that negative and positive outliers are asymmetric: there are considerably more positive outliers and they also have different powers in asymptotic behaviour. We model this observations with a distribution with mentioned properties, a non central Student`s t-distribution, which can cover asymmetric 
non-Gaussian probability distributions. Noise probability function is then given by:
\begin{equation} \label{eq:pdfNoise}
 p(n) =  (1 - a) \, N(\mu,\sigma) \, + \, a \; NCT(\frac{n - b}{c}, \nu, \psi)  ,
\end{equation}
where NCT is a non central t-distribution with parameters $\nu$ and $\psi$. $b$ and $c$ are used to rescale $n$. Parameter $a$ is a measure of the relative impact of outliers compared to the Gaussian contribution and is on the order of one percent.
Power law decay of Student`s distribution for large values dominates Gaussian distribution while small $a$ make it negligible for small $y$.

Optimal parameters of the Gaussian and non-Gaussian 
components of analytic pdf and their relative weight $a$ can be extracted by minimizing negative log-likelihood of the star flux data with no planets. The resulting noise model is shown in Figure \ref{fig:noise_model}, together with the data. We obtain 
a satisfactory fit to the data.

\section{Flux variability model} \label{chap:Star}

To develop the flux variability model 
we assume two components, one for star's variability and another for planet transits. In this work we ignore additional 
possible components, such as eclipsing binaries. 
We will then proceed in the next section to perform joint likelihood analysis of the two, allowing for a combination of planets and star variability, given the non-Gaussian noise distribution encoded by 
the likelihood. 

\subsection{Star variability}

We approach star variability as an non-parametric Gaussian process whose hyperprior we determine from the data. Gaussian 
process assumes the data are Gaussian distributed around 
the mean, with 
correlations between the flux values. These
correlations help determine the star flux as a function of 
time, and it is crucial that the correlations are properly
described. Typically in a Gaussian process one uses a kernel description of the correlation function, 
with a simple form such a Gaussian or Mattern kernel, 
and a few parameters only to describe the kernel. If correlation structure is complex such a description is not sufficient. Stellar variability 
models can be very complex, exhibiting phenomena such as quasi-periodic oscillations on several scales \cite{star_variability}.
We expect the process to be stationary, and 
for this reason 
we will use Fourier basis expansion, in which case the 
Gaussian hyperprior is the power spectrum of the flux. 
This can be modeled essentially non-parametrically, by 
evaluating the power as a function of frequency in 
terms of bandpowers, where several nearby Fourier modes are 
averaged over, such that the resulting power spectrum estimate 
has sufficiently small error. 
Once 
this is determined, the flux reconstruction is minimal 
variance (and hence optimal), 
and goes under the name of Wiener filter. Our model 
differs from other Gaussian process approaches such as 
celerite \cite{celerite}, in that our kernel optimization 
is essentially non-parametric and thus close to optimal. 
It is also very fast, since it is based on Fast Fourier 
Transforms (FFT). We apply small amount of zero padding 
at the ends to avoid issues with periodic boundary 
conditions with FFTs, treating them as masked data, 
similar to how we treat gaps in the data. Zero padding
and gaps lead to 
a power suppression that can be easily corrected for with 
simulations. 

We work with stellar variability in the frequency domain. Fourier components of star`s flux are $s(\nu) = \mathscr{F}^{-1}(s(t))= u(\nu) + i v(\nu)$, satisfying $s(-\nu) = s(\nu)^{\dagger} = u(\nu) - i v(\nu)$, as we have real valued signal in the time domain and complex 
Fourier modes. The goal of this optimization is to find optimal $s(\nu)$ by minimizing negative log-likelihood function. We have a hyperprior, the power spectrum $P(\nu) = \langle |s(\nu)|^2 \rangle$, 
that needs to be determined first. We will 
assume that stellar variability is a Gaussian
process, which we will verify by the final result of the analysis. 

\subsection{Planet transition}
We will model planetary transition with two parameters determining limb darkening profile of a star ($u_1,\,u_2$), which will be considered known in the process, and four parameters of interest which are properties of a planet: period ($T$), phase ($\phi$), depth of transit ($A$) and time of transit ($\tau$). Shape of transit is primarily determined by a star`s flux density $j$, that is intensity per surface area of a cross section, emitted in the direction of sight. Limb darkening makes it a function an angle $\alpha$ determined by observer, center of a star and point on the surface of a star emitting flux. It will be approximated by a second order polynomial in $\cos{\alpha}$ as proposed in \cite{simple_limb_darkening}:
\begin{equation}
\frac{j(\cos{\alpha}; u_1, u_2)}{j(1; u_1, u_2)} = 1 - u_1 (1-\cos{\alpha}) - u_2 (1-\cos{\alpha})^2.    
\end{equation}
More sophisticated models exist \cite{complicated_limb_darkening}, but we found they are not required for our purpose here, and the 
model we use has sufficiently small residuals already. We integrate over a planet`s shadow to get a total flux reduction.

relWe assume constant velocity of a planet during the time of transit and neglect orbital eccentricity. Orbital inclination is taken into account by leaving time of transition as a parameter, and not fixing it to the $\tau = R_*\sqrt[3]{4 T / \pi G M_*}$, as would be expected for a perfectly aligned transit using Kepler`s law. Radius of a planet ($r$) is not an independent parameter but can be calculated from $A$:
\begin{equation}
    A = \frac{ j(1; u_1, u_2) \, \pi (\frac{r}{R_*})^2}{2 \pi \int j(\cos{\alpha}; u_1, u_2) \cos{\alpha} \, \textit{d} \cos{\alpha}}= \frac{(r/R_*)^2}{1-u_1/3 - u_2/6},
\end{equation} 
where $R_*$ is the radius of a star. It impacts flux profile through integral over a planet`s shadow and in the edge of the transit where only a part of a planet`s shadow covers a star. Both effects are small when radius is sufficiently small and will be computed from the initial guess on a planet`s radius and held fixed during the optimization. If this assumption is not valid, as for example for binary stars, optimization should be iterated with respect to the radius. For our purposes it is valid and enables us to prepare a shape of transit in advance and represent it with a spline interpolation $\mathcal{S}(t \, ; \, r,\, u_1,\, u_2)$. We take into account finite time lapse between measurements $\Delta t$, so that what we measure is in fact an average value of flux during this time lapse. Thus the shape of transit for $t \in [0, T)$ is given by
\begin{equation}\label{eq:Ushape}
    U(t, T, \phi, A, \tau) = \frac{A}{\Delta t} \int_{t-\frac{\Delta t}{2}}^{t+\frac{\Delta t}{2}} \mathcal{S}\bigg(\frac{t' - \phi}{\tau}  \, ; \, r,\, u_1,\, u_2\bigg) \text{d}t',
\end{equation}
which in the absence of time transit variations (TTV) 
repeats periodically with a period $T$.

\subsection{Joint analysis}\label{chap:Joint}

\begin{figure*}[t]
\includegraphics[scale = 0.45]{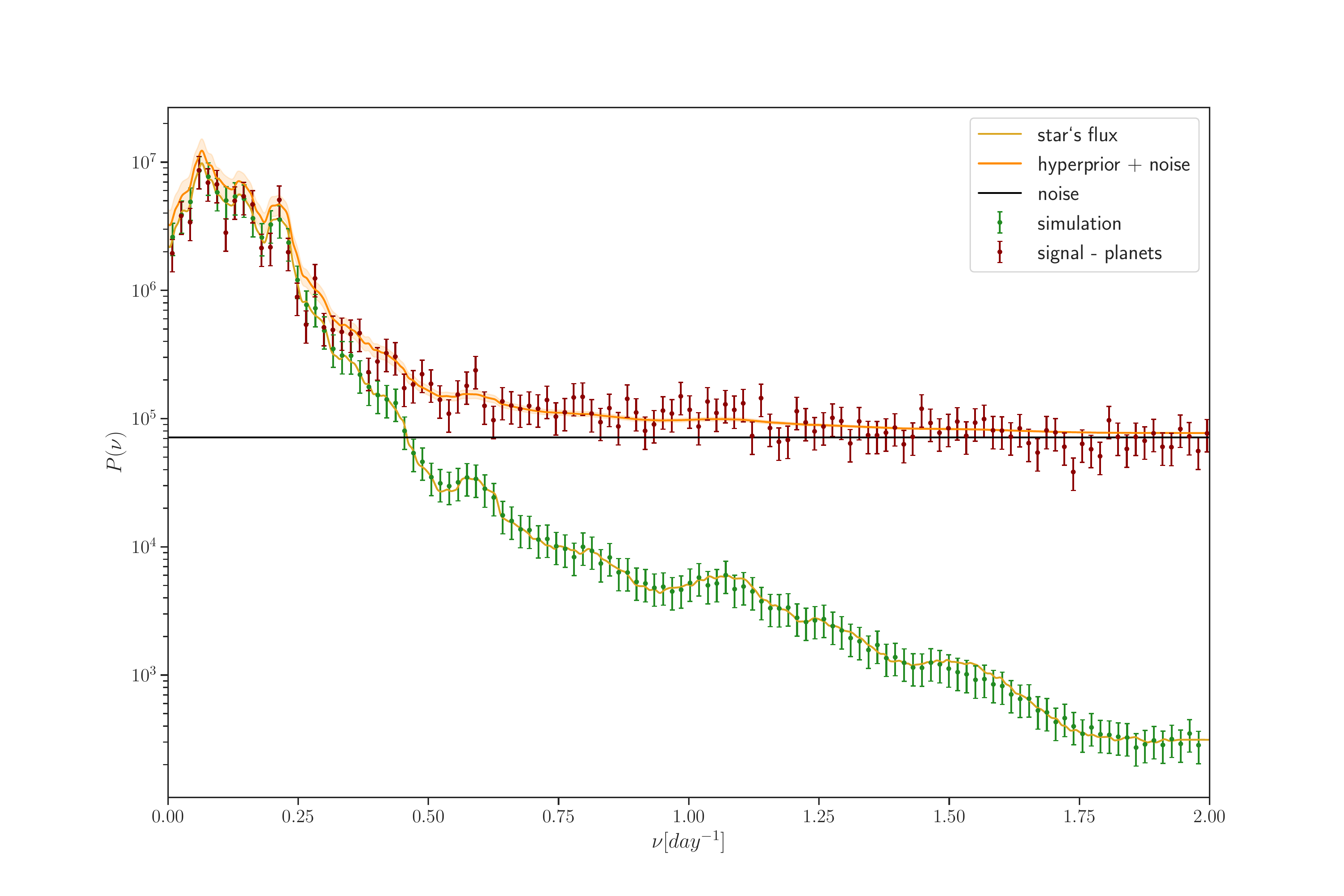}
\caption{Power spectrum of Kepler-90 (without planets):
Wiener filter signal (star's flux), as well as  
star variability true power spectrum (hyperprior+noise), 
with one sigma error 
confidence band after the iterations. Flux signal from 
planets is eliminated by joint analysis. A sign of convergence is that the power spectrum drawn from a simulation with the given hyperprior
results in a simulated Wiener filter that coincides with star`s Wiener filter. At high frequencies Wiener filter gives significantly lower power than true power from 
hyperprior, a consequence of power being lower than noise power, which drives Wiener filter to zero. We see that the 
power spectrum contains low frequency power  caused
by star variability. There is a 25\% offset between hyperprior + noise and true signal power due to gaps in the Kepler-90 
time ordered data, where data are zeroed.}
\label{fig:hyperprior}
\end{figure*}

In our model the observed data $y(t)$ is composed of three terms: star variability, planet transitions and noise, so that the 
noise is 
\begin{equation}
   n(t)= y(t) -  s(t) - \sum_{i} U(t, T_i, \phi_i, A_i, \tau_i) .
   \label{nt}
\end{equation}
where $U$ are shapes of transits defined in equation \ref{eq:Ushape}, $T_i$, $\phi_i$, $A_i$, $\tau_i$ are parameters of i\textsuperscript{th} planet, $s(t)$ is stellar flux and $n(t)$ is a realization of noise distributed according to equation \ref{eq:pdfNoise}. In previous chapters we developed models for each contribution. 
In order to extract parameters of planet transitions and confidence that they are indeed planets, not just fluctuations in stars' flux, both star variability and planets should be analyzed simultaneously.
\\
\\
Posterior of the model parameters is the product of the data likelihood $p(n)$
and prior. The former is given by the noise 
probability distribution of equation \ref{eq:pdfNoise}, while the latter is given 
by the Gaussian probability distribution of $p(s(\nu))$, with variance
given by the power
spectrum $P(\nu)$: we assume flat non-informative 
prior for planet terms. 
We minimize the loss function $\mathcal{L}$, which is the 
negative log posterior against $s(\nu)$ and $T_i$, $\phi_i$, $A_i$, $\tau_i$, iterating on the power spectrum
$P(\nu)$,
\begin{equation}
\mathcal{L}= -2\sum_t \ln p(n(t)) + \sum_\nu \left[\frac{|s(\nu)|^2}{P(\nu)}+\ln P(\nu)\right],
\end{equation}
where $n(t)$ is given by equation \ref{nt} and $p(n(t))$ by equation 
\ref{eq:pdfNoise}. The first sum 
is over all the measured time stream data: if there is 
no data in certain time bands we simply omit it from the 
sum. 

Minimization of $\mathcal{L}$ can be performed in high dimensional space of planet parameters and Fourier modes. 
It is aided by availability of gradient for FFT modes, so 
despite many components (typically thousands of modes), 
optimization is not expensive. On the other hand, planet parameters are only a few, but it is inefficient to compute the gradient at each step. As a result, joint optimization is slow, and a better approach is to iterate optimization with respect to Fourier components at fixed planets parameters, and with respect to planets parameters while holding Fourier components fixed. That is, we first eliminate impact of planets with current best guess and find optimal Fourier modes, then we fix star variability and find a better guess on planet parameters, and we keep iterating until convergence. Joint analysis gives us parameters of planets and star Fourier modes in a few iterations, given a hyperprior on a spectral power of a star and initial guess on parameters of planets, which are both unknown. 
\\
\\
\subsection{Power spectrum hyperprior estimation}

We would like to determine the hyperprior power spectrum 
$P(\nu)$ to be as parameter free as possible, while at the 
same time being determined with a sufficiently small error, 
so that the subsequent Gaussian process/Wiener filter analysis is reliable. Star variability among 
different stars is very diverse, 
so using other stars is unlikely to be useful to 
determine the hyperprior, and for this reason we will 
determine it only from the data of the star itself. 
In the absence of noise each positive 
Fourier frequency $\nu$ has 
two components, and the relative error on the power 
spectrum is $\delta P(\nu)/P(\nu)=2/2=1$. 
By combining $N$ nearby frequencies into 
a bandpower we can reduce the error to $\delta P(\nu)/P(\nu)=N^{-1/2}$. Noise further increases the 
error: if noise is white with power spectrum amplitude 
$P_n$ then $\delta P(\nu)/P(\nu)=(1+P_n/P(\nu))N^{-1/2}$. 
Here we will choose the width of the bandpower such that 
the relative 
error is approximately constant and of order 0.2, which 
means averaging over 25 frequencies at low frequencies
where noise is negligible, and more at high frequencies
where noise is large compared to the signal. An error 
of 0.2 at each bin ensures that overfitting $P(\nu)$
is not important for reconstruction, 
as the effect of a fluctuation 
of this order on the Wiener filter reconstruction is small.

Initial guess for a hyperprior power spectrum can be any function that resembles a power spectrum of a star: we use a flat hyperprior at low frequencies and a power law decay at high frequencies. In the n\textsuperscript{th} step of iteration joint fit gets us optimal Fourier modes of a star $s^{(n)} (\nu)$ given our current best guess on a hyperprior $P^{(n)}$. We simulate a power spectrum of a star and noise under the assumption that $P^{(n)}$ is already a correct hyperprior. We choose random phases with given power $P^{(n)}$, transform it in a time domain, add noise drawn from noise probability distribution model found in \autoref{chap:Noise}, add zeros when Kepler did not measure flux and in the zero padding region at the edges, transform back in the frequency domain and compute power. We repeat this random realization of $P_{sim}$ to compute an expected value $\langle P_{sim} \rangle$. We compare simulated power spectrum with the optimal power spectrum of a star $|s^{(n)}|^2$ found from the joint fit assuming hyperprior $P^{(n)}$. Hyperprior in the next step of iteration is then:
\begin{equation}
    P^{(n+1)}(\nu) = P^{(n)}(\nu) \; + \; |s^{(n)}|^2 (\nu)\;-\; \langle P_{sim}^{(n)}(\nu) \rangle.
\end{equation}
High frequency variations in the hyperprior (frequency in the frequency domain) are a consequence of noise, and we  eliminate it by a low pass filter before doing a next step of iteration. Hyperprior converges in a few iterations. Simulated power spectrum reproduces the 
true star variability spectrum, as seen in 
figure \ref{fig:hyperprior}, which is a sign that we 
have converged onto the correct hyperprior.


\section{Application to planet detections} \label{chap:planet}

We have developed a method for simultaneously fitting star variability, noise and planet parameters. In this paper we will assume that planet candidates with initial guesses on parameters are known: our goal is not a planet search method, but 
improving the statistical analysis of planetary parameters such as transit period and phase, and planet radius. 
For this reason we apply it in this section 
to simulations, and to 
the known planets 
of Kepler 90. 

\subsection{Comparison on simulations}

We first test our method on simulated flux, comparing our method with a method widely used in literature, showing it is more accurate in reproducing planet transit amplitude. We then compare both methods on a concrete example of a light curve of Kepler 90. 
\\
\\
A method frequently used in literature is fitting spline to eliminate star variation, and then separately fit planets. To prevent over fitting, spline method is repeatedly applied with different knot spacings to obtain the lowest Bayesian information criterion coefficient \cite{spline}. In contrast, in our joint fit method the hyperprior converges to the optimal power across the frequency range, with no need to use heuristics such as Bayesian information criterion \cite{BICcritique}. Another advantage of our 
joint fit approach is that it does not eliminate any flux that could contain information about planets, which results in a better reproduction of planet`s transit amplitude and better signal to noise. 

\begin{figure*}[t]
\includegraphics[scale =0.38]{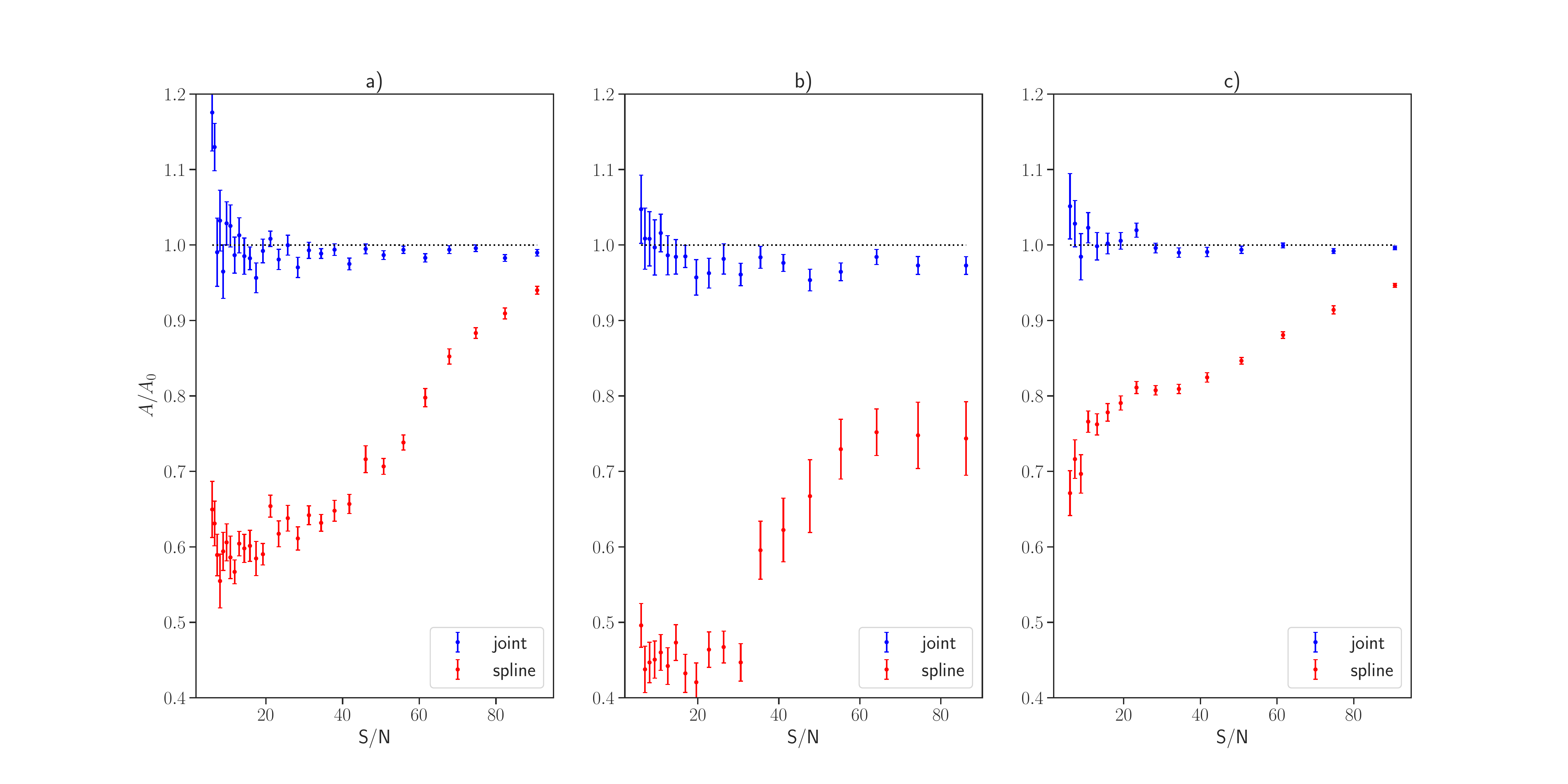}
\caption{Amplitude obtained by the fit is compared to the known injected value as a function of S/N of a simulated planet for both methods. a) At each S/N of a planet simulation with uniformly distributed period in the range 7-100 days and uniformly distributed phase. Amplitude $A$ is averaged over the realizations of the noise, star variability and planet parameters. b) Star variability is taken to be fixed and the same as Kepler 90 star variability. Amplitude is averaged over the noise realizations and planet parameters. c) Star variability is fixed to Kepler 90 star variability and period is fixed to be equal to Kepler 90 c planet (8.7 days). Amplitude is averaged only over planet phase. Fixing the period reduces variance of the amplitude, a sign that amplitude bias is period dependent. In all cases joint fit with non-Gaussian noise and Gaussian process star variability is almost unbiased estimator of amplitude and planet`s radius (since $A \propto r^2$), while spline fit results in significant bias of the 
amplitude.}
\label{fig:test}
\end{figure*}

A performance comparison of both methods is shown in figure \ref{fig:test}. We simulate star variability (random phases of  Fourier components with power resembling power spectrum of a Kepler 90), white noise, and a planet transit with parameters typical for small inner planets (10 days period and 4 hours time of transition). We vary the amplitude $A$ of planet`s transit depth, which is directly proportional to the signal to noise ($S/N$) of a planet. We compare both methods in their ability to reproduce the simulated amplitude. As shown in figure \ref{fig:test}, 
joint fit results in a more reliable estimate of amplitude, 
that is within 2\% of correct values, except for low S/N, 
where it rises to 5\% deviation at S/N=6. In contrast, 
the corresponding spline fit is systematically too low, 
with the negative bias reaching 40\% at S/N<40. 

\subsection{Kepler 90 analysis}

We will now demonstrate these methods on a concrete example of a star Kepler-90. We chose this star because it is known to host seven or eight planets and has received a lot of attention in the literature. We used light curves that were pre-processed by Kepler pipeline. This includes bias voltage correction, calibrated pixels, removed artifacts like cosmic rays and identified pixels of target stars \cite{kepler_process}. 
We use long cadence light curves, which is a 1460 days long signal (with gaps), with 29.4 minutes spacing. We normalize flux in different quarters as described in \ref{chap:Noise}, subtract average and add zeros with infinite variance in places where there is no data so quarters can be concatenated in one signal with evenly spaced measurements, unit variance and zero average. 
Starting from published planet parameters we 
do a joint fit of planets and star variability. 

\begin{figure*}[t]
\includegraphics[scale = 0.5]{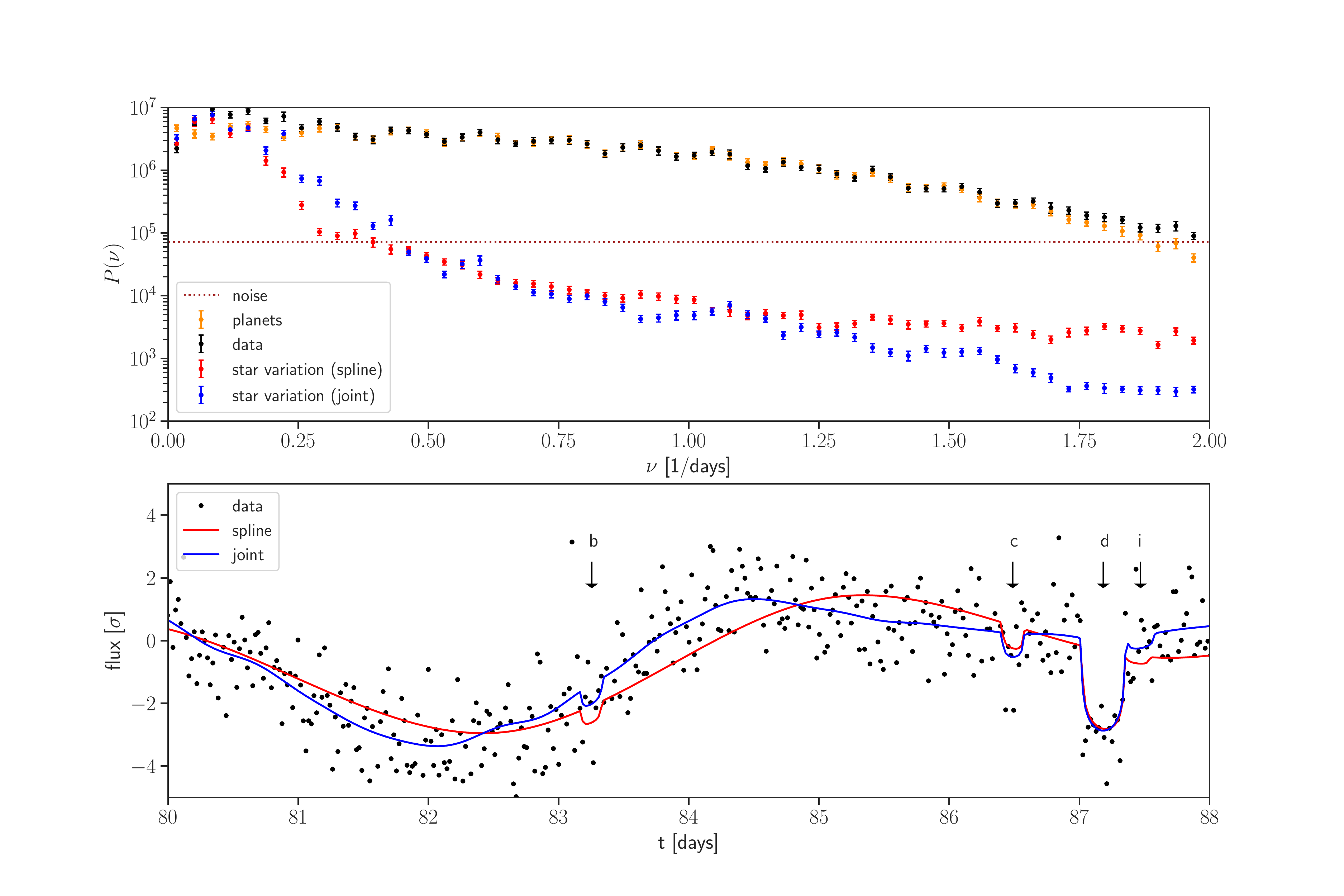}
\caption{a) Top panel:
power spectrum of planet signal and star variability 
for Kepler-90 without any planet removal. 
Most of spectral power at high frequencies comes from U shape of planet transitions and only insignificant contribution comes from star variability. Joint fit isolates planets in the data, allowing determination of the star variability 
power spectrum. 
b) Bottom panel:
separate spline fit and planets` fit (red), a standard procedure in literature, is compared to joint fit of Fourier components and planets (blue). 
One can observe that the joint fit is better at fitting star variability, while at the same time improving the fit to the planets, 
resulting in deeper U shaped fits. Time is measured relative to the beginning of Kepler's measurements in Kepler-90 (HJD-2454833 + 131.5124).}
\label{fig:comparisson}
\end{figure*}

\begin{table*}[t]
\caption{A comparison of signal to noise ratios and radius of planets in Kepler 90 system using different methods. Only 
statistical error is included in the joint fit radius, while 
the official radius error also includes the contribution from the error on the star radius.}
\begin{tabular}{|l|c|c|c|c|c|c|c|c|}\hline%
\label{table}
\bfseries planet & \bfseries spline & \bfseries spline & \bfseries joint fit & \bfseries official &\bfseries spline &\bfseries joint & \bfseries official\\
\bfseries  & \bfseries $\sqrt{\Delta \chi^2}$ & \bfseries $\sqrt{-2 \Delta \ln p}$ & \bfseries $\sqrt{-2 \Delta \ln p}$ & \bfseries $S/N$ & \bfseries $r [R_{\oplus}]$ & \bfseries $r [R_{\oplus}]$ & \bfseries $r [R_{\oplus}]$
\begin{filecontents*}{deltalog.csv}
planet,splineGauss,splineP,myR,splineR,officialR,myP,official
b,16.2,16.8,1.29 $\pm$ 0.04,1.22,1.31 $\pm$ 0.17,15.5,16.7
c,14.0,15.5,1.36 $\pm$ 0.05,1.22,1.19 $\pm$ 0.14,15.8,16.6
i,4.9,5.4,1.02 $\pm$ 0.06,0.81,1.3 $\pm$ 0.2,6.7,unknown
d,35.1,35.1,2.83 $\pm$ 0.03,2.63,2.9 $\pm$ 0.3,29.6,18.9
e,28.6,28.6,2.72 $\pm$ 0.04,2.49,2.7 $\pm$ 0.3,23.5,25.1
f,24.6,24.6,2.69 $\pm$ 0.06,2.7,2.9 $\pm$ 0.5,16.6,30.2
g,224.1,61.6,7.72 $\pm$ 0.02,7.72,8.1 $\pm$ 0.8,113.4,121.8
h,336.6,49.9,10.82 $\pm$ 0.02,10.81,11.3 $\pm$ 1.0,142.6,233.3
\end{filecontents*}
\csvreader[head to column names]{deltalog.csv}{}%
{\\ \hline \planet & \splineGauss & \splineP & \myP & \official & \splineR & \myR & \officialR}%
\\\hline
\end{tabular}
\end{table*}

We compare our method in time and in frequency domain, to show our method is better at fitting planets and star variability, as shown in Figure \ref{fig:comparisson}.  
Top panel shows that planets dominate the spectral density
at high frequencies, a consequence of strong 
planetary signatures in this system. 


Results for seven known planets and a proposed eight planet Kepler-90 i are given in table \ref{table}. In the first two columns, spline is repeatedly fitted to find optimal Bayesian information criterion, eliminated, and then planets are fitted. An absence of individual planet in the fit gives a higher $\chi^2$, and we report the square root of the difference. If we take in account non Gaussian contribution of noise we report the corresponding value of square root of $-2 \Delta \ln p$. Third column shows $[-2 \Delta \ln p]^{1/2}$ for the joint analysis with Fourier nodes and planets fitted together. The fourth column is from NASA Exoplanet Archive \cite{officialsnr}. It is a systematic search through many solar systems and does not take in account out of phase transits of Kepler-90g and Kepler-90 h, which results in a lower SNR for these planets, compared to the spline fit using Gaussian noise. We can see that joint fit often gives lower signal to noise when compared to the spline fit, 
despite the fact that we predict higher signal to noise
on simulated planets. We argue this is 
a consequence of improved fit of star variability, and 
that our model is better at modeling star as a source of false positives. 
Kepler-90 g and Kepler-90 h (two biggest planets) have lower $-2 \Delta \ln p$ than $\Delta \chi^2$. Their large amplitudes in $\chi^2$ are likely to be noise fluctuations caused by outliers.
\\
\\
Perhaps more importantly, 
it is evident from table \ref{table} that signal to noise ratio can be dramatically changed if star and planets are fitted together and if true noise probability distribution 
is used instead of Gaussian assumption. This is a sign that outlier contribution to the noise probability distribution is important and that simplification to do a separate fit of planets and star variability is not justified. 
\\
\\
Fifth and sixth columns show that spline fitting method generally underestimates radius of a planet, which is observed in even greater significance when averaged over many realizations of the signal simulations, as shown in the figure \ref{fig:test}. The last column is an official estimate on planet`s radius \cite{officialR}. One can observe that 
official errors are significantly larger than those found
from our analysis, but this is because we only include 
the statistical error in our joint analysis, while official 
error also includes the error on the star radius, which is 
uncertain at about 10\% level, while we used a fixed value of 1.2
solar radius in our fits. 

In terms of the estimated radius we observe
discrepancies between spline and joint fits, which 
are consistent with figure \ref{fig:test}. We chose 
to explicitly investigate it further 
for Kepler-90 c. In figure 
\ref{fig:test} we show results of a synthetic 
analysis of such a 
planet, where we inject it into the data at a random phase 
with a period of 8.7 days and with an amplitude 
consistent with Kepler-90c, and then analyze it with 
the two methods. Results show that spline 
analysis underestimates the radius by about 10\%, comparable 
to the difference we observe between spline fit and 
joint fit of actual Kepler-90c. 

\section{Conclusions}

In this paper we develop methodology for analyzing stationary time 
series data, such as star flux data, in the presence of 
non-Gaussian noise. After the data have been preprocessed 
to account for calibrations
one is left with a calibrated time series, which may have non-Gaussian noise distribution. As a first step 
we develop likelihood analysis using the true noise probability 
distribution. In the presence of outliers this 
distribution reduces their impact on the fit, and makes the 
analysis robust without the need to eliminate outliers 
by hand. In this paper we define noise as the difference
between the observed data and the signal that combines
star variability and planet transits. As a consequence, 
the noise probability distribution can only be determined
as part of a joint iterative analysis that also determines the star 
variability and planet transits as the same time. 

Next we address the star variability using Gaussian 
processes. We adopt non-parametric hyperprior using 
Fourier space power spectrum, and develop an iterative 
procedure that determines the power spectrum together
with the star variability reconstruction, which is a 
Wiener filter of the data given the power spectrum. 
In contrast to existing methods \cite{spline}, ours has the 
advantage of having more flexibility in terms of 
the hyperprior, which is essentially non-parametric and 
can fit even detailed features in the star spectrum. 
It is also very fast 
since it is based on Fast Fourier Transforms. 

We apply our method to simulated planets and show
that we recover the signal more accurately than the 
current practice of a separate spline fit, followed by a
planet transit fit. We also apply our method to 
real data of Kepler-90, and show that it 
gives considerably different results when compared to 
splines with Gaussian and non-Gaussian error 
distribution, as well as when compared to official 
signal to noise and radius numbers. This shows that the 
specifics of analysis can affect the results, and 
while we only analyzed the effects on the transit amplitude, 
we expect there will be similar effects on the 
other planetary parameters, such as period, phase or TTVs. We argue that our approach 
is close to optimal and whenever high precision planetary 
transit parameters are needed, a joint star variability and planet transit analysis should be performed, together 
with the proper noise non-Gaussian likelihood analysis.
All of these aspects are included in our current version 
of the code, which is freely available\footnote{\url{https://github.com/JakobRobnik/Kepler-data-analysis}}. 
Other 
astrophysical sources can be added to the joint fit, such as eclipsing binaries, which we plan to do in the future.



\acknowledgements We thank Stephen Bryson for useful discussions. We acknowledge ASEF for providing funding for JR visit to UC Berkeley. This material is based upon work supported by the National Science Foundation under Grant Numbers 1814370 and NSF 1839217, and by NASA under Grant Number 80NSSC18K1274.
\bibliographystyle{aasjournal.bst}
\bibliography{cosmo,citations}

\end{document}